\documentclass[aps,prb,preprint]{revtex4}
\usepackage{graphicx}
\usepackage{latexsym}
\usepackage{amsmath}
\usepackage{amssymb}
\usepackage{float}
\usepackage{epstopdf}
\usepackage{bm}

\begin{document}
\baselineskip=24pt
\title{Exploring Foundations of Time-Independent Density Functional 
Theory for Excited-States}
\author{Prasanjit Samal}
\email{Fax:+91-512-259 0914 ; E-mail: dilu@iitk.ac.in}
\author{Manoj K. Harbola}
\email{Fax:+91-512-259 0914 ; E-mail: mkh@iitk.ac.in}
\affiliation{Department of Physics, Indian Institute of Technology,
 Kanpur U.P. 208016, India}
\begin{abstract}
Based on the work of G\"{o}rling and that of Levy and Nagy, density-functional 
formalism for many Fermionic excited-states is explored through a careful and 
rigorous analysis of the excited-state density to external potential mapping. 
It is shown that the knowledge of the ground-state density is a must to fix 
the mapping from an excited-state density to the external potential. This 
is the excited-state counterpart of the Hohenberg-Kohn theorem, where instead of 
the ground-state density the density of the excited-state gives the true many-body 
wavefunctions of the system. Further, the excited-state Kohn-Sham system is
defined by comparing it's non-interacting kinetic energy with the true kinetic
energy. The theory is demonstrated by studying a large number of atomic systems. 
\end{abstract}
\maketitle
\newpage
\section{INTRODUCTION}
Hohenberg-Kohn-Sham(HKS) density functional theory(DFT) \cite{hk,ks,jg} has been 
most widely used to investigate the electronic structures of many-electron 
systems. It is a theory for dealing with the ground states and their properties 
\cite{py,dg,nhm} . Applications of the Hohenberg-Kohn theorem and the Kohn-Sham 
construction is limited to the ground-state because it is the ground-state 
density of an electronic system that determines the Hamiltonian, and consequently 
other physical observables of the electronic system. This suggests both the ground 
as well as excited-state properties can be determined from the ground-state density 
through the Hamiltonian operator since it characterizes all the states of a system. 
On the other hand, to develop an excited-state DFT akin to it's ground-state
counterpart, it is important to describe an excited-state of a system in terms of 
the density of that state. Almost for the last two and half decade researchers 
have investigated \cite{gl,zrb,vb,th,pathak,efh,pl,gok,sen,nagy,sd,ng1} the possibility of 
giving a formal foundation to the time-independent excited-state DFT like the HKS 
DFT does for the ground-states, but it is yet to come to its full fruition.\\  

To make the excited-state calculations feasible within the density functional
formalism there are two open questions to be answered: (i) Does there exist
a mapping between the excited-state density and the external potential like
the ground-states ? (ii) Secondly, for the determination of the excited-
state energies, is it possible to construct reasonably accurate 
exchange-correlation energy functionals? We note that although the exact form of the
exchange-correlation functional is unknown for the ground-states, there are 
several accurate and approximate functionals \cite{b88,pw91,lyp,gga,mgga}in traditional
DFT. The issue is to find such approximate functionals for the excited-states. The second 
question is partly answered through an attempt made by the present authors\cite{psmkh} in 
the recent past by developing an exchange energy functional within the local-density 
approximation(LDA) for a particular class of excited-states. This suggests that 
a correlation functional can also be developed in a similar fashion.\\

We now address the first question which is the main focus of the present study. Over 
the past few years, a lot of attention has been paid to the question of mapping 
\cite{ag,ln,smss,harb,gb,sh} from an excited-state density $\rho(\vec r)$ to the external 
potential $v_{ext}(\vec r)$, because the entire structure of time-independent excited-state 
DFT depends on that. A brief account of this is as follows :\\

The first step in establishing a mapping from an excited-state density to a many
electron wavefunction is taken by looking for $\rho$-stationary states \cite{ag}.
These are states $\Psi$ that reproduce a given density $\rho_k$(density of the $k^{th}$-
excited-state) and simultaneously make the expectation value $\left<\Psi|\hat T + \hat V_{ee}
|\Psi\right>$ stationary. However, for a given density there are many $\rho$-stationary 
states and thus establishing a mapping requires an additional input. Levy and Nagy 
\cite{ln,ng1} provide this by requiring that $\Psi$ be orthogonal to $\Psi_j(j < k)$, which 
are to be determined by the ground-state density $\rho_0$ for the system under study. Thus 
the knowledge of the ground-state density is essential to define a map 
$\rho_k\longrightarrow\Psi_k$. This is reasonable because it is the ground-state density 
that really fixes the Hamiltonian of a system uniquely. Levy and Nagy then go on to 
construct a Kohn-Sham system of non-interacting electrons whose $m^{th}$ excited-state 
produces the given excited-state density. Further they put forth a criterion that the 
ground-state density of the Kohn-Sham system is closest to the true ground state density 
of the system in the least square sense. \\

We have been studying the mapping from an excited-state density to the potential
and have been investigating the Levy-Nagy criterion for constructing the Kohn-Sham
system for excited-states. Our work in this direction forms the content of this 
paper. We show:\\
(i) an explicit construction of the external potential from an excited-state density 
using G\"{o}rling 's approach \cite{ag},\\
(ii) that the Levy-Nagy criterion \cite{ln,ng1} of constructing the Kohn-Sham system
is not perfect. We give reasons for it,\\
(iii) that for a given state the Kohn-Sham system should be constructed by comparison 
of the kinetic energies of the true and the non-interacting systems.

\section{G\"{o}rling/Levy-Nagy Formalism}
\subsection{G\"{o}rling's Stationary-State Formulation}
We start with G\"{o}rling's formulation \cite{ag} of the excited-state problem. The
formulation is based on the constrained-search approach \cite{cs} and provides a map
from an excited-state density to a many-body wavefunction.  For a given 
excited-state density $\rho(\vec r)$ a $\rho-$stationary state $\Psi(\vec r)$ is given by 
making the expectation value $\left<\Psi|\hat T + \hat V_{ee}|\Psi\right>$ stationary   
with the constraint that the many-particle wavefunction $\Psi(\vec r)$ giving the
same density $\rho(\vec r)$ . Corresponding to each $\rho-$stationary state $\Psi(\vec r)$ 
there is an external potential $v_{ext}(\vec r)$ . This has been shown by G\"{o}rling. 
We give a different proof of it here, with the external potential $v_{ext}(\vec r)$ arising 
as the Lagrange multiplier to ensure the constraint of producing the density $\rho(\vec r)$.

A $\rho-$stationary wavefunction $\Psi$ gives, by the stationarity principle
\begin{equation}
\left<\delta\Psi|\hat T + \hat V_{ee}|\Psi\right> + 
\left<\Psi|\hat T + \hat V_{ee}|\delta\Psi\right> = 0
\end{equation}
with the constraint that
\begin{equation}
\delta\rho(\vec r) = \int\left\{\Psi\delta\Psi^\dagger + 
\Psi^\dagger\delta\Psi\right\}{d\vec r_2}.....{d\vec r_N} = 0
\end{equation}
Writting Eq.~{1} in expanded form, we get
\begin{equation}
\int\left\{\delta\Psi^\dagger\left(\hat T + \hat V_{ee}\right)\Psi
+ \Psi^\dagger\left(\hat T + \hat V_{ee}\right)\delta\Psi\right\}
{d\vec r_1}{d\vec r_2}.....{d\vec r_N} = 0
\end{equation}
Because of condition (Eq.~{2}) above this will be satisfied if  
\begin{equation}
\int \left(\hat T + \hat V_{ee}\right)\Psi(\vec r_1,\vec r_2,...,\vec r_N){d\vec r_2}
.....{d\vec r_N} = \int f_1(\vec r_1)\Psi(\vec r_1,\vec r_2,...,\vec r_N){d\vec r_2}
.....{d\vec r_N}\;.
\end{equation}

Thus in general $\rho-$stationarity of $\Psi$ implies that it satisfies 
\begin{equation}
\left(\hat T + \hat V_{ee}\right)\Psi(\vec r_1,\vec r_2,...,\vec r_N) = \left(\sum_i 
f_i(\vec r_i)\right)\Psi(\vec r_1,\vec r_2,...,\vec r_N)\;.
\end{equation}

However, since $\hat T, \hat V_{ee}$ are symmetric operators and $\Psi(\vec r_1,\vec r_2
,...,\vec r_N)$ is antisymmetric, it's necessary $\left(\sum_i f_i(\vec r_i)\right)$ must 
also be symmetric. Thus all $f_i$'s must be the same function $f(\vec r)$. Identifying this 
function as $f(\vec r) = -v(\vec r) + E$ , where $\raisebox{-1.6ex}{$\stackrel{\displaystyle
{\text{lim}}}{\scriptstyle{\vec r\to +\infty}}$}v(\vec r) = 0 $, we get $\Psi$ satisfying
\begin{equation}
\left\{\hat T + \hat V_{ee} + \sum_i v(r_i) \right\}\Psi(\vec r_1,\vec r_2,...,\vec r_N)
= E \Psi(\vec r_1,\vec r_2,...,\vec r_N) \,.
\end{equation}  
We note that for different $\rho-$stationary states $v_{ext}(\vec r)$ will be different.
Thus by applying the constrained search method we get many $\rho-$stationary
states and the corresponding external potentials. The question is which one of these 
corresponds to a given system. Levy and Nagy identify \cite{ln,ng1} this system as the 
one where $\Psi$ is orthogonal to $\Psi_j(j<k)$ for a given ground-state density 
$\rho_0$. Thus in the Levy-Nagy theory \cite{ln,ng1}, the wavefunction $\Psi[\rho;\rho_0]$ 
is a bi-functional of $\rho$ and $\rho_0$. One subtle point about the Levy-Nagy theory is 
that if the search for $\Psi$ is restricted to the space orthogonal to $\Psi_j(j<k)$, the 
variational principle becomes a minimum principle. The prescription above also makes the 
functional 
$F[\rho;\rho_0] \; = \; \raisebox{-1.6ex}{$\stackrel{\displaystyle{\text{min}}}
{\scriptstyle{\Psi \rightarrow \boldsymbol{\rho}({\bf r}) }}$}
\langle \Psi | \hat{T} \;+\; \hat{V}_{ee} |\Psi \rangle $ 
a bi-functional of the excited-state density $\rho(\vec r)$ as well as the ground-state 
density $\rho_0(\vec r)$ .\\

\begin{figure}[thb]
\includegraphics[width=5.0in,height=3.3in,angle=0.0]{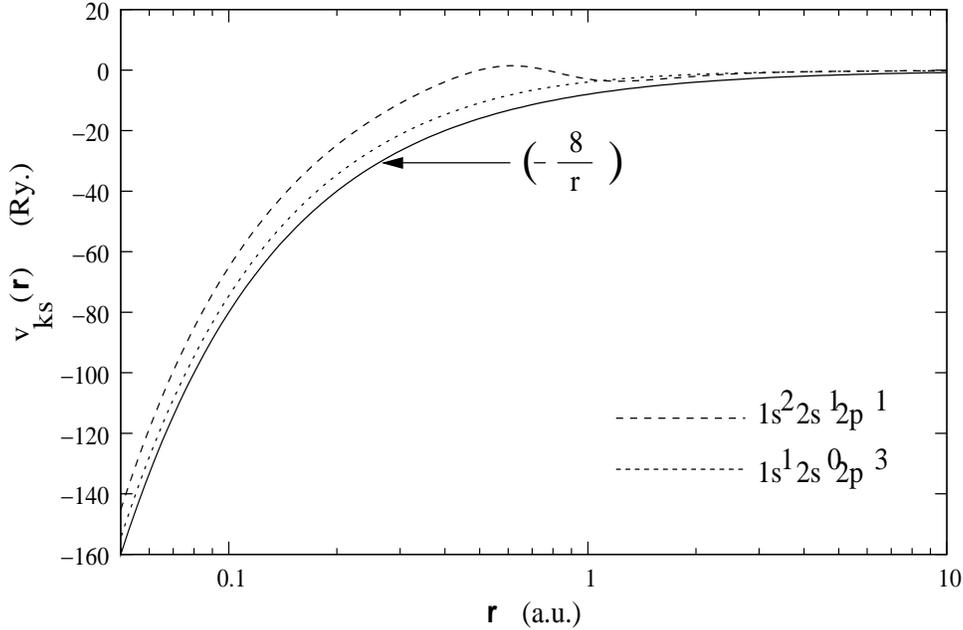}
\caption{Shown in the figure are the KS potentials for the excited-state density of
$1s^1 2s^0 2p^3~(^5{S})$ state of $Be$ generated for the original and one alternative 
configuration.}
\label{pot-comp}
\end{figure}

Next we discuss how a Kohn-Sham(KS) system can be realized for an excited-state density.
To get a KS system, the mapping from a given excited-state density to a non-interacting
wavefunction is established by making the expectation value $\left<\Phi|\hat T|\Phi\right>$
stationary with respect to variations in $\Phi$ with $\Phi$ giving the excited-state
density. How this is done is described later. As is the case for the interacting 
systems, there may be several $\Phi$'s and the corresponding KS potentials $v_{KS}$ 
that give rise to an excited-state density and make $\left<\Phi|\hat T|\Phi\right>$ 
stationary. Two such potentials for the density of $1s^1 2s^0 2p^3~(^5{S})$ of $Be$ 
are shown in Fig.~{1} . The excited-state density used is obtained by solving the 
excited-state KS equation with the Harbola-Sahni(HS) \cite{hs} exchange-only potential. 
The question that arises again is how do we choose one particular KS system to represent 
a system in its excited-state. An intuitive way would be by comparing the ground-state 
densities, as was done for the interacting systems.  However, the ground-state density 
of a non-interacting system that reproduces an excited-state density may not be the same 
as that of the true system (it will be the same only if the electron-electron interaction 
were absent). Thus Levy and Nagy propose \cite{ln,ng1} that of the many Fermionic 
non-interacting systems that give the same excited-state density, the one whose ground-state 
density resembles the exact one in the least square sense be identified as the Kohn-Sham 
system. The criterion is obviously exact \cite{sh} for systems with no electron-electron 
interaction, as stated above. For interacting-electron systems, the criterion appears to 
give \cite{sh} the true Kohn-Sham system, consistent with the orbitals to which
the electrons have been excited, as we discuss below.   

\subsection{Levy-Nagy Criterion}
We have been investigating the Levy-Nagy criterion for many different excited-states of
atomic systems. We find that whereas the criterion is satisfied for a large number of 
excited-states, there are exceptions also. We present and discuss these results below. 
The results are obtained as follows. Working within the central field approximation, 
we perform the excited-state calculations using the exchange-only HS method and obtain 
the excited-state density (KLI-OPM \cite{kli,opm,ng2}  gives similar results). We then 
use the Zhao-Morrison-Parr(ZMP) method \cite{zmp} to generate the same density with 
different configurations and test the Levy-Nagy criterion for these configurations. 
In the ZMP method the Kohn-Sham system for a given density $\rho(\vec r)$ and a 
configuration of choice \cite{harb} is obtained by solving the equation
\begin{equation}
\left\{-\frac{1}{2}\nabla^2 + \lambda\int\frac{\left[\sum_j n_j|\phi_j(\vec r\prime)|^2 - 
\rho(\vec r\prime)\right]}{|\vec r - \vec r\prime|} d\vec r\prime \right\} \phi_i(\vec r) 
= \varepsilon_i\phi_i(\vec r)
\end{equation}
in the limit of ${\lambda\to \infty}$ . Here $\left\{n_j\right\}$ are the occupation numbers 
of the orbitals according to the configuration chosen. We choose $\lambda$ large enough so 
that not only do the densities match to a high degree of accuracy, the highest occupied 
eigenvalues $\varepsilon_{max}$ of the above equation also matches with the original 
$\varepsilon_{max}$ to within $5 \%$; in fact it is better than within $2 \%$ in many of the
cases. For example, for the $1s^1 2s^0 2p^3$ excited-state of $Be$, we have generated the 
same density with three other configurations: $1s^2 2s^0 2p^2 , 1s^2 2s^1 2p^1$ and 
$1s^1 2s^1 2p^2$. The $\varepsilon_{max}$ and the expectation values 
$\left<\frac{1}{R}\right>,\left<R\right>$ and $\left<{R}^2\right>$ for 
different configurations are compared in Table~{I}. To judge the numerical accuracy of our 
ZMP program, we also generate the excited-state density with the original configuration and 
compare numbers obtained with the original numbers. We see that with the original 
configuration, the $\varepsilon_{max}$ comes to within $2 \%$ of the original value with 
$\lambda = 5000$ whereas the various expectation values are essentially exact. For the 
three alternative configurations, the accuracy of 
$\left<\frac{1}{R}\right>,\left<R\right>$ and $\left<{R}^2\right>$ 
is about the same but the $\varepsilon_{max}$  values differ slightly more depending on the 
configuration. The worst case is the $1s^1 2s^1 2p^2$ for which 
$\varepsilon_{max} = -0.626~Ry$ for $\lambda = 30000$. To make sure that the eigenvalue will
eventually converge to $\varepsilon_ {max} = -0.658~Ry$, we performed calculations for 
different values of $\lambda$ for this configuration and found that 
$\varepsilon_{max} = -0.621~Ry , -0.624~Ry$ and $-0.626~Ry$ for $\lambda = 10000 , 14000$ 
and $30000$, respectively, thereby shifting towards the true eigenvalue albeit very slowly. 
We also mention that for the configuration $1s^2 2s^1 2p^1$, the uppermost orbital 
is $2s$ and not $2p$. The local potential in which the electrons are move is then given as: 
\begin{equation}
v_{KS}(\vec r) = \lambda\int \frac{\left[\sum_j n_j|\phi_j(\vec r\prime)
|^2 - \rho(\vec r\prime)\right]}{|\vec r - \vec r\prime|} d\vec r\prime \;.
\end{equation}
Having found the potential above, we obtain the ground-state density $\tilde\rho_0(\vec r)$ 
of this potential by occupying the lowest energy orbitals with the given number of electrons.
We calculate its mean square distance from the true ground-state density $\rho_0(\vec r)$ as
\cite{sh}
\begin{equation}
\Delta\left[{\rho_0(\bf r)},\tilde{\rho}_0{(\bf r)}\right] = 
\int_\infty\left\{{\rho_0(\bf r)} - \tilde{\rho}_0{(\bf r)}\right\}^2 {\rm d}^3 r \;.
\end{equation}
By the true ground-state density here, we mean the ground-state density obtained with the 
Harbola-Sahni exchange potential. The results for $\Delta$ for a number of atoms and their 
excited-states are shown in Tables~{II} $\&$ {III} . The original configuration of the 
excited-sates is shown in the second column. The third column shows the alternative 
configurations using which we obtain the same density and the fourth column the 
corresponding $\Delta\left[{\rho_0(\bf r)},\tilde{\rho}_0{(\bf r)}\right]$. It is seen from 
the results that for most of the cases $\Delta$ is the smallest for the original 
configuration but there are cases where $\Delta$ is smallest for a different configuration. 
For example, there is the excited-state $1s^1 2s^0 2p^3$ of $Be$ for which the 
configuration $1s^1 2s^1 2p^2$ gives the smallest $\Delta$. 
Similarly for the state $1s^2 2s^2 2p^3 3s^2$ of $F$,  $1s^2 2s^2 2p^6 3s^2$ of $Ne$ and 
$1s^1 2s^0 2p^3$ of $B^+$, $1s^1 2s^0 2p^3$ of $Ne^{6+}$ $\Delta$ is the smallest for a 
configuration other than the original configuration of the system. Thus we find that the 
Levy-Nagy criterion, as quantified by Eq.~{8} above, is not satisfactory in that it leads 
to a KS system where an excited-state configuration is not consistent with the original system.

\subsection{An Alternative Criterion}
Having found that the Levy-Nagy criterion is not fully satisfactory to identify a KS system,
we have looked for other ways of doing so, remaining within the Levy-Nagy proposal of
comparing the ground-state densities. Thus instead of comparing densities directly, we
compare them energetically as follows. After obtaining many different non-interacting 
systems for an excited-state density, we take their ground-states and calculate the 
expectation value of the true ground-state Kohn-Sham Hamiltonian(constructed using the 
HS exchange potential) with respect to these ground-states. Thus the 
calculation proceeds as follows: we solve the HS equation for the ground-state of a system 
and obtain the ground-state Kohn-Sham Hamiltonian $H_0$. The expectation value 
$\left<H_0\right>$ with respect to the true ground-state orbitals is designated as 
$\left<H_0\right>_{true}$; it is the sum of the eigenvalues of the ground-state orbitals. 
Next we take the different non-interacting systems giving an excited-state density, 
consider their ground-states and calculate the expectation value $\left<H_0\right>_{alt.}$ 
with respect to these states. Because of the variational principle 
$\left<H_0\right>_{alt.}$ should always be above $\left<H_0\right>_{true}$. We then identify
the true KS system as that for which $\left<H_0\right>_{alt.}$ is closest to 
$\left<H_0\right>_{true}$. This comparison is made in Tables~{IV $\&$ V} for the same 
systems as in Tables~{II and III} with HS density. As is clear from the table the alternate 
criterion is better in that the correspondence between the original system and the Kohn-Sham
system is restored for $F$,$Ne$ and $Ne^{6+}$. However, new inconsistencies arise in $Al$,$Si$ 
and $P^+$ although in these cases the difference in the numbers for the original and the
alternative configuration is very small. On the other hand, the inconsistency in $Be$ and 
$B^+$ remains. We note that this criterion is very sensitive 
to the exchange potential. If calculations are done with KLI-OPM exchange potential, the 
inconsistency remains only in $Be$ and $B^+$ systems. It is clear from the discussion in 
the two sections above that a criterion based on comparison of ground-state densities of 
excited-state Kohn-Sham systems cannot be satisfactory.

\section{Present Theory}
Given the background above, we now present a consistent theory of excited-states within 
the rubric of density-functional approach. The principal tenets of the theory are:

({\bf i}) There is a straightforward way of mapping an excited-state density $\rho(\vec r)$
to the corresponding many-electron wavefunction $\Psi(\vec r)$ or the external potential
$v_{ext}(\vec r)$ using the $\rho$-stationary wavefunctions. The wavefunction depends 
upon the ground-state density $\rho_0$ implicitly.

({\bf ii}) The Kohn-Sham system is defined through a comparison of the kinetic energy
for the excited-states. This avoids any comparison of the ground-state densities which, 
as seen above, doesn't give a satisfactory way of defining a KS system.
We now discuss these two points one by one.

To describe the mapping from an excited-state density $\rho_k(\vec r)$ to a many-body
wavefunction, we take recourse to the constrained-search approach. This gives, as
discussed earlier, many different wavefunctions $\Psi_k(\vec r)$ and the corresponding
external potential $v^k_{ext}(\vec r)$. If in addition to the excited-state density we also
know the ground-state density $\rho_0$  then $v_{ext}(\vec r)$ is uniquely determined 
by the Hohenberg-Kohn \cite{hk} theorem. Thus with the knowledge of $\rho_0$ it is quite 
straightforward to select a particular $\Psi$ that belongs to a $\left[\rho,\rho_0\right]$ 
combination by comparing $v^k_{ext}(\vec r)$ with $v_{ext}(\vec r)$. Alternatively, we can 
think of it as finding $\Psi$ variationally for a $\left[\rho,v_{ext}\right]$ combination
because the knowledge of $\rho_0$ and $v_{ext}$ is equivalent. Through the constrained 
search above a functional
\begin{equation}
F[\rho;\rho_0] \; = \;
\langle \Psi[\rho;\rho_{0}] | \hat{T} \;+\; \hat{V}_{ee} | \Psi[\rho;\rho_{0}]\rangle 
\label{constrained_search}
\end{equation}
is also defined. The prescription above is similar to that of Levy and Nagy \cite{ln}
but avoids the orthogonality condition imposed by them. \\

The densities for different excited state for a given ground-state density $\rho_0$ or 
external potential $v_{ext}$ can thus be found as follows: take a density and search for 
$\Psi$ that makes $\left<\Psi|\hat T + v_{ee}|\Psi\right>$ stationary; check whether 
the corresponding $v_{ext}$ matches with the given $\rho_0$(or $v_{ext}$); if not, take 
another density and repeat the procedure until the correct $\rho$ is found. Also because of 
the proof given in section {II} above, the Euler equation for the excited-state density is 
\begin{equation}
\frac{\delta F\left[\rho,\rho_0\right]}{\delta\rho(\vec r)} + v_{ext}(\vec r) = \mu
\end{equation} 
where $\mu$ is the Lagrange multiplier to ensure that $\rho_k(\vec r)$ integrates to
the proper number of electrons.

The prescription above for the excited-states in terms of their densities is quite
straightforward, particularly because it's development is parallel to that for the
ground-states. On the other hand, to construct a Kohn-Sham \cite{ks} system for
a given density is non-trivial; and to carry out accurate calculations for excited-
states it is of prime importance to construct a KS system. Further, a KS system
will be meaningful if the orbitals involved in an excitation match with the 
corresponding excitations in the true system. We have shown above that the Kohn-Sham 
system constructed using the Levy-Nagy criterion fails in this regard.

In principle, obtaining a Kohn-Sham system is quite easy. Define the non-interacting
kinetic energy $T_s\left[\rho,\rho_0\right]$ and use it to further define the 
exchange-correlation functional as 
\begin{equation}
E_{xc}\left[\rho,\rho_0\right] = F\left[\rho,\rho_0\right] - \frac{1}{2}\int\int
\frac{\rho(\vec r)\rho(\vec r\prime)}{|\vec r - \vec r\prime|} d\vec r d\vec r\prime 
- T_s\left[\rho,\rho_0\right]\;.
\end{equation}
Then the Euler equation for the excited-state densities will read
\begin{equation}
\frac{\delta T_s\left[\rho,\rho_0\right]}{\delta\rho(\vec r)} +
\int\frac{\rho(\vec r\prime)}{|\vec r - \vec r\prime|} d\vec r\prime
+ \frac{\delta E_{xc}\left[\rho,\rho_0\right]}{\delta\rho(\vec r)} + v_{ext} = \mu\;.
\end{equation}
which is equivalent to solving 
\begin{equation}
\left\{- \frac{1}{2}\nabla^2 + v_{KS}(\vec r) \right\}\phi_i(\vec r) = \varepsilon_i\phi_i(\vec r)
\end{equation}
with
\begin{equation}
v_{KS}(\vec r) = v_{ext}(\vec r) + \int\frac{\rho(\vec r\prime)}{|\vec r - \vec r\prime|} 
d\vec r\prime + \frac{\delta E_{xc}\left[\rho,\rho_0\right]}{\delta\rho(\vec r)}\;.
\end{equation}
However, it is defining $T_s\left[\rho,\rho_0\right]$ that is not easy in the 
excited-state problem. For the ground-states, $T_s\left[\rho_0\right]$ is easily
defined as the minimum kinetic energy for a given density obtained by occupying the lowest
energy orbitals for a non-interacting system. On the other hand, for the excited-states 
it is not clear which orbitals to occupy for a given density, particularly because 
a density can be generated by many different non-interacting systems. Levy-Nagy select 
one of these systems by comparing the ground-state density of the excited-state
non-interacting system with the true ground-state density. However, this criterion
is not satisfactory as discussed earlier. Therefore some other criterion has to be
evolved to construct the excited-state Kohn-Sham system.\\ 

Before searching for other ways of constructing a Kohn-Sham system, we look for reasons
that may be responsible for the Levy-Nagy criterion not being fully satisfactory. We 
argue that we are not being consistent while comparing the ground-state density of an 
excited-state KS system with the true ground-state density. This is because the 
ground-state density of the excited-state KS system is not the self-consistent
ground-state density of the $v_{ext}(\vec r)$ obtained for the excited-state
density but of a potential different from $v_{ext}(\vec r)$ . In comparing the ground-state
densities, we are thus not comparing the $v_{ext}(\vec r)$ of the excited-state KS system 
with the true $v_{ext}(\vec r)$, and this at times leads to inconsistent results.\\

In light of the above remarks, it is important that in constructing the Kohn-Sham system 
only the self-consistently determined quantities corresponding to a given excited-state
density be compared. Thus we propose that the
KS system be so chosen that it is energetically very close to the original system.
To ensure this, we define the KS system as that system for which the non-interacting
kinetic energy , obtained through constrained search over the non-interacting 
wavefunctions, is closest to $T\left[\rho,\rho_0\right]$, already obtained through the 
full constrained search. 
This then gives the functional $T_s\left[\rho,\rho_0\right]$. Thus $T_s\left[\rho,\rho_0
\right]$ is defined as the kinetic energy that is closest to the true kinetic energy 
$T\left[\rho,\rho_0\right]$ obtained for a given excited-state density $\rho$. Defining 
$T_s\left[\rho,\rho_0\right]$ in this manner also keeps the DFT exchange-correlation
energy close to the conventional quantum mechanical exchange-correlation energy. An added 
advantage of keeping the difference between the two
kinetic energies $T_s\left[\rho,\rho_0\right]$ and $T\left[\rho,\rho_0\right]$ 
smallest is that the structure of the Kohn-Sham potential is simple; it is known 
that contribution of $\underbrace {T - T_s}$ gives more structure to the KS
potential.\\

All the statements in this paragraph are justified on the basis of the differential virial
theorem \cite{hm}. Using this theorem the exchange-correlation potential for a given density
and the corresponding many body wavefunction $\Psi$ can be written as
\begin{equation}
-\nabla v_{xc}(\vec r) = \frac{\left\{ \vec Z_{KS}(\vec r;\left[{\Gamma_1}^{KS}\right]) - 
\vec Z(\vec r;\left[{\Gamma_1}\right]) + \int\left [\nabla u(\vec r,\vec r\prime)\right] 
\left[ \rho(\vec r)\rho(\vec r\prime) - 2\Gamma_2(\vec r,\vec r\prime)\right]d\vec r\prime
\right\}}{\rho(\vec r)} 
\end{equation}
where $\Gamma_{1}(\Gamma_2)$ is the first(second) order density matrix and the vector 
fields $\vec Z,\vec Z_{KS}$ are related to the true and Kohn-Sham kinetic energy density 
tensors respectively. For a given
$[\rho,\rho_0]$, $\vec Z,\Gamma_{1},\Gamma_2$ are fixed and different configurations chosen give
different $\vec Z_{KS}$. These different configurations can be thought of as arising 
from a different external potential (as is shown below) or from a different 
exchange-correlation potential \cite{smss,harb}. In any case kinetic energy difference 
between the true and Kohn-Sham system is \cite{vs}
\begin{equation}
\Delta T = \frac{1}{2}\int \vec r \mathbf{.} \left\{\vec Z_{KS}(\vec r;\left[{\Gamma_1}^{KS}
\right]) - \vec Z(\vec r;\left[{\Gamma_1}\right])\right\} d\vec r\;.
\end{equation}
It is this difference that we propose be kept the smallest for the true KS system,
and as we show below, it gives the Kohn-Sham system consistent with the original system.

\subsection{Examples}
We now demonstrate the ideas presented above with examples. Since we do not know how
to perform the general constrained search $\rho\rightarrow\Psi,\delta\left<\Psi|\hat T
+ \hat v_{ee}|\Psi\right> = 0$, we take an indirect path for the purpose of demonstration.
In the following we work with atomic excited-state densities generated by HS
exchange-only potential and take these as the excited-state density.  The densities and 
the energies obtained by the HS formalism are essentially the same as those of 
Hartree-Fock(HF)\cite{fischer} theory for both the ground as well as the excited-states. 
Similarly the HS exchange potential is very close to the true local exchange potential of 
the optimized potential method(OPM). Thus the formalism is well suited to test the ideas 
presented above. Thus we start with this given $[\rho,v_{ext}]$ combination.\\

\begin{figure}[thb]
\includegraphics[width=5.0in,height=3.3in,angle=0.0]{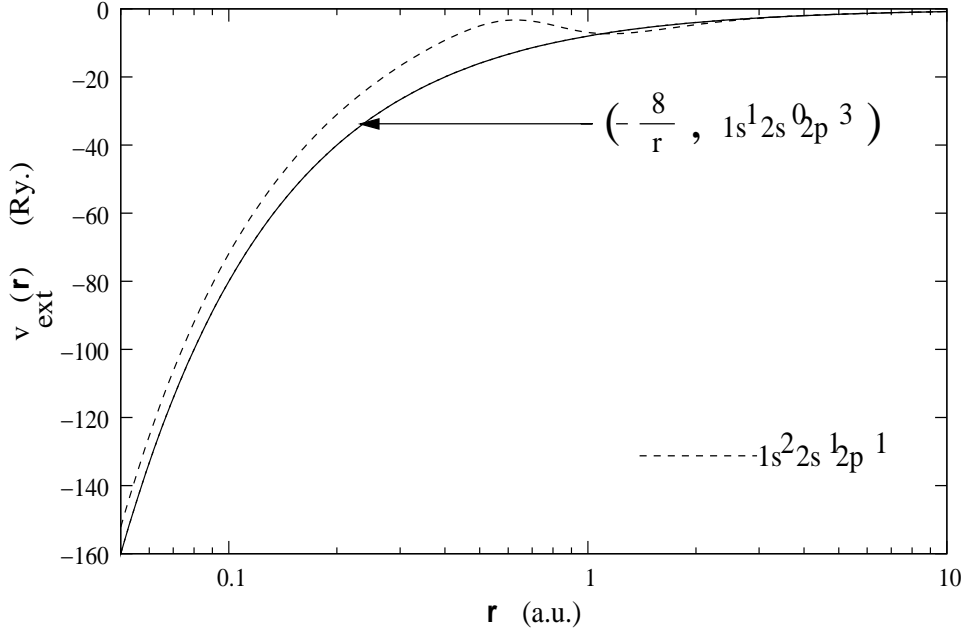}
\caption{Shown in the figure are the external potentials for the excited-state density 
of $1s^1 2s^0 2p^3~(^5{S})$ state of $Be$ corresponding to the original and one alternative
configuration.}
\label{pot-comp1}
\end{figure}

We first demonstrate that an excited-state density is generated by different external
potentials depending on the configurations chosen to generate the density. For this we
use the ZMP method to get the non-interacting potential giving the density $\rho(\vec r)$ 
and subtract from it the Coulomb $v_{coul}(\vec r)$ and the exchange ${v_x^{HS}}(\vec r)$ 
potential to get the external potential $v_{ext}(\vec r)$ ({\em i.e.}~$v_{ext}(\vec r) = 
v_{KS}(\vec r) - v_{coul}(\vec r) - v_{x}^{HS}(\vec r)$) . The exchange potential for a 
given set of occupied orbitals is obtained using the Harbola-Sahni formula \cite{zope} for 
it. Shown in Fig.~{2} are different external 
potentials thus generated for the $1s^1 2s^0 2p^3~(^5{S})$ of density $Be$. To check our 
consistency, we first obtain $v_{ext}(\vec r)$ for the original configuration 
($1s^1 2s^0 2p^3$) and find it correctly to be $- \frac{8}{r}~ Ry$ . The other 
configurations that we use to obtain the same density are 
$1s^2 2s^0 2p^2 , 1s^2 2s^1 2p^1 , 1s^1 2s^1 2p^2$. We have shown only two potentials 
corresponding to the configurations $1s^2 2s^1 2p^1$ and $1s^1 2s^0 2p^3$, and compared them
with the true external potential $v_{ext}=- \frac{8}{r}~ Ry$.  As discussed earlier, only 
one of these - that corresponding to the original configuration - matches with the 
true external potential.\\

\begin{figure}[thb]
\includegraphics[width=5.0in,height=3.3in,angle=0.0]{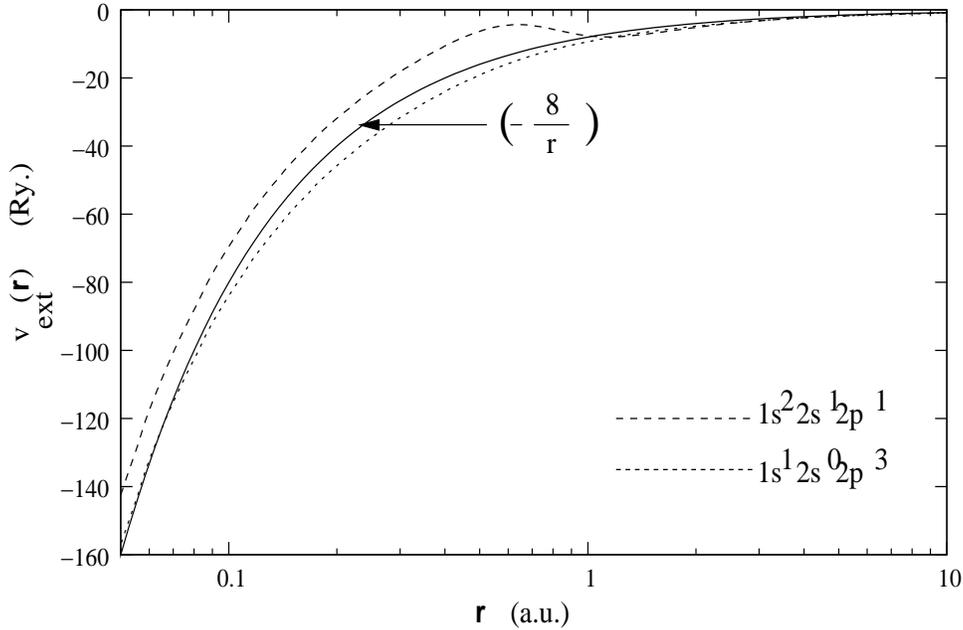}
\caption{Shown in the figure are the external potentials corresponding to the ground-state 
densities of excited-state KS systems for $1s^1 2s^0 2p^3~(^5{S})$ state of $Be$.  The
potentials are compared with the true external potential $v_{ext}=- \frac{8}{r}$. }
\label{pot-comp2}
\end{figure}

Next we show that the external potentials corresponding to the ground-state density of
excited-state KS systems are different from the external potential for the 
excited-state density. This, as pointed out earlier, sometimes leads to 
non-satisfaction of the Levy-Nagy criterion. Shown in Fig.~{3} are the 
${\tilde v}_{ext}(\vec r)$ corresponding to the ground-state densities of different 
configurations for the excited-state density of $1s^1 2s^0 2p^3~(^5{S})$ state of $Be$.
These potential are also obtained by subtracting from the Kohn-Sham potential $v_{KS}$
the Coulomb and the Harbola-Sahni exchange potential calculated by occupying the 
corresponding KS orbitals in the ground-state configuration.  We see that whereas the 
true external potential is $-\frac{8}{r}~ Ry$, the external potentials corresponding to
the ground-states are different.  It is this difference that leads to the ground-state
densities different from the true one, and also sometimes to inconsistencies between the 
KS and the true systems.\\

Finally we show that a comparison of the kinetic-energies(KE) leads to the appropriate 
Kohn-Sham system. We demonstrate this with the densities generated using the HS exchange
potential and with one example with correlated density. The numbers for the noninteracting
kinetic energy for different configurations corresponding to the same excited-state 
densities as considered earlier are shown in Tables~{VI} and {VII} and are compared with 
their original kinetic energy.  Since the HS potential is a local potential itself, the 
correct configuration gives $\Delta T_s = 0$ (slight difference arising due to numerical 
calculations) and wrong ones a larger value, as is evident from the numbers in the Tables. 
We see that unlike the Levy-Nagy criterion, a comparison of excited-state 
KE leads to a proper KS system in $\underline{all}$ the cases. Of course when we use the 
correlated densities, the difference $\Delta T $ is not going to be $zero$ for the proper 
configuration but should be the smallest. This is clearly seen in the example of 
$1s 2s (^1{S})$ state of He atom where the true $T[\rho]$ is $2.146~a.u.$. We have used the 
ZMP procedure to obtain the KS potentials in this case also and see that KE for the
$1s^2$ and the $1s2s $ configurations is $2.044~a.u.$and $2.153~a.u.$, respectively. In the 
latter case it is closer to the true kinetic energy of the system. Thus the configuration 
$1s2s$ represents the KS system for the $1s 2s (^1{S})~He$ density.\\

\section{Concluding Remarks}
Building on the work of G\"{o}rling and Levy and Nagy, we have presented in this
paper a consistent theory of excited states within the density-functional formalism.
The theory is based on constrained-search and defines a bi-density functional
$F[\rho,\rho_0]$ without the orthogonality constraint of Levy-Nagy. Further the
theory gives a clear definition of the excited-state Kohn-Sham systems as that
whose kinetic and exchange-correlation energy components are closest to those of the 
true system. This avoids the problem of comparing the non-self-consistent ground-state
densities, as proposed in the LN theory, so no inconsistency arises in identifying 
an excited-state Kohn-Sham system.\\

To conclude, we have analyzed the theoretical foundations of excited-state 
time-independent density-functional theory and have put Kohn-Sham calculations for
excited-states on a firm footing.  It is clear from our present work that an excited-state 
configuration of the KS system corresponds to a similar excited-state of the true system 
(with the major component of the excited-state wavefunction involving the same orbitals as 
the KS system).  The work should also help in providing guidance in construction of the 
excited-state exchange-correlation energy functionals to facilitate self-consistent 
determination of the excited-state energies.

\newpage

\begin{table}
\caption{Shown in the table $\varepsilon_{max}$ and expectation values $\left<\frac{1}{R}\right>
,\left<R\right> \& \left<{R}^2\right>$ for various configurations giving the same density
as that of $1s^1 2s^0 2p^3; ^5{S}$ state of $Be$. The self-consistently determined values
of these physical quantities are : $\varepsilon_{max} = -0.658~Ry , \left<\frac{1}{R}\right> = 
5.818 , \left<R\right> = 6.755 \& \left<{R}^2\right> = 17.309$.}
\vspace{0.2in}
\begin{tabular}{|c|c|c|c|c|c|}
\hline
configurations&$\lambda$&$\varepsilon_{max}(Ry.)$&$\left<\frac{1}{R}\right>$&$\left<R\right>$
&$\left<{R}^2\right>$\\
\hline
$1s^1 2s^0 2p^3$&5000&-0.649&5.818&6.755&17.312\\
\hline
$1s^2 2s^0 2p^2$&5000&-0.649&5.819&6.755&17.312\\
\hline
$1s^2 2s^1 2p^1$&5000&-0.655&5.819&6.755&17.312\\
\hline
$1s^1 2s^1 2p^2$&30,000&-0.626&5.818&6.755&17.310\\
\hline

\end{tabular}
\end{table}

\begin{table}
\caption{Value of $\Delta$(See Eq.~{8}) for different configurations($3^{rd}$ column) giving
the same excited-state density as that for the original configuration($2^{nd}$ column).
Systems where the LN criterion is not satisfactory are indicated with a `$\ast$' against
them.}
\vspace{0.2in}
\begin{tabular}{|c|c|c|c|}
\hline
atoms/ions &$true. config.$& $alt. config.$ &  $\Delta\left[{\rho_0(\bf r)},\tilde{\rho}_0{(\bf r)}\right]$ \\
\hline
& &$1s^1 2s^1 2p^2$&0.0662\\
$^{\ast}Be$&${1s^1 2s^0 2p^3}$&$1s^1 2s^0 2p^3$&0.1627\\
& &${1s^2 2s^1 2p^1}$&0.8758\\
\hline
& &$1s^2 2s^0 2p^3$&0.0002\\
$B$&${1s^2 2s^0 2p^3}$&${1s^2 2s^1 2p^2}$&0.0065\\
& &$1s^2 2s^2 2p^1$&0.0286\\
\hline
& &$1s^2 2s^1 2p^3$&0.0008\\
$C$&${1s^2 2s^1 2p^3}$&${1s^2 2s^2 2p^2}$&0.0156\\
& &$1s^2 2s^0 2p^4$&0.0181\\
\hline
& &$1s^1 2s^0 2p^6$&0.2903\\
$N$&${1s^1 2s^0 2p^6}$&${1s^2 2s^0 2p^5}$&6.8461\\
& &$1s^2 2s^1 2p^4$&8.8409\\
\hline
& &$1s^1 2s^1 2p^6$&0.3875\\
$O$&${1s^1 2s^1 2p^6}$&${1s^2 2s^0 2p^6}$&8.9609\\
& &$1s^2 2s^1 2p^5$&12.1269\\
\hline
& &$1s^2 2s^1 2p^6$&0.0002\\
$F$&${1s^2 2s^1 2p^6}$&${1s^2 2s^2 2p^5}$&0.0621\\
\hline
& &${1s^2 2s^2 2p^4 3s^1}$&0.0521\\
$^{\ast}F$&${1s^2 2s^2 2p^3 3s^2}$&$1s^2 2s^2 2p^3 3s^2$&0.2704\\
\hline
& &${1s^1 2s^2 2p^6 3s^1}$&0.7138\\
$^{\ast}Ne$&${1s^1 2s^1 2p^6 3s^2}$&$1s^1 2s^1 2p^6 3s^2$&1.6176\\
\hline
\end{tabular}
\end{table}

\begin{table}
\caption{Caption is the same as in Table~{II}.}
\vspace{0.2in}
\begin{tabular}{|c|c|c|c|}
\hline
atoms/ions &$true. config.$& $alt. config.$ &  $\Delta\left[{\rho_0(\bf r)},\tilde{\rho}_0{(\bf r)}
\right]$ \\
\hline
& &${1s^1 2s^1 2p^2}$&0.0402\\
$^{\ast}B^+$&${1s^1 2s^0 2p^3}$&$1s^1 2s^0 2p^3$&0.1931\\
& &$1s^2 2s^1 2p^1$&2.0261\\
\hline
 & &${1s^1 2s^1 2p^2}$&0.1465\\
$^{\ast}Ne^{6+}$&${1s^1 2s^0 2p^3}$&$1s^1 2s^0 2p^3$&0.3463\\
\hline
& &$1s^2 2s^2 2p^6 3s^0 3p^2$&0.0007\\
$Mg$&${1s^2 2s^2 2p^6 3s^0 3p^2}$&${1s^2 2s^2 2p^6 3s^1 3p^1}$&0.0014\\
& &$1s^2 2s^2 2p^6 3s^2 3p^0$&0.0039\\
\hline
 & &$1s^2 2s^2 2p^6 3s^0 3p^3$&0.0008\\
$Al$&${1s^2 2s^2 2p^6 3s^0 3p^3}$&${1s^2 2s^2 2p^6 3s^1 3p^2}$&0.0022\\
 & &$1s^2 2s^2 2p^6 3s^2 3p^1$&0.0073\\
\hline
 & &$1s^2 2s^2 2p^6 3s^1 3p^3$&0.0007\\
$Si$&${1s^2 2s^2 2p^6 3s^1 3p^3}$&${1s^2 2s^2 2p^6 3s^2 3p^2}$&0.0027\\
 & &$1s^2 2s^2 2p^6 3s^0 3p^4$&0.0078\\
\hline
 & &$1s^2 2s^2 2p^6 3s^0 3p^3$&0.0004\\
$Si^{+}$&${1s^2 2s^2 2p^6 3s^0 3p^3}$&${1s^2 2s^2 2p^6 3s^1 3p^2}$&0.0042\\
 & &$1s^2 2s^2 2p^6 3s^2 3p^1$&0.0148\\
\hline
 & &$1s^2 2s^2 2p^6 3s^1 3p^3$&0.0005\\
$P^{+}$&${1s^2 2s^2 2p^6 3s^1 3p^3}$&${1s^2 2s^2 2p^6 3s^2 3p^2}$&0.0053\\
 & &$1s^2 2s^2 2p^6 3s^0 3p^4$&0.0099\\
\hline
 & &$1s^2 2s^2 2p^6 3s^0 3p^5$& 0.0006\\
$P$&${1s^2 2s^2 2p^6 3s^0 3p^5}$&${1s^2 2s^2 2p^6 3s^1 3p^4}$&0.0055\\
 & &$1s^2 2s^2 2p^6 3s^2 3p^3$&0.0207\\
\hline
\end{tabular}
\end{table}

\begin{table}
\caption{Comparison of the expectation value $-\langle H_0\rangle_{alt.}$ and $-\langle H_0
\rangle_{exact}$ for various configurations corresponding to a given excited-state density.
The first column mentioned are the atoms/ions, the second column the original configuration
, fourth column the alternative configurations considered. In the third and last column given
the expectations values $-\langle H_0\rangle_{exact}$ and $-\langle H_0\rangle_{alt.}$
respectively. }
\vspace{0.2in}
\begin{tabular}{|c|c|c|c|c|}
\hline
atoms/ions & $exact~config.$ & $-\langle H_0\rangle_{exact}~{\bf a.u.}$&$alt.~config.$&
$-\langle H_0\rangle_{alt.}~{\bf a.u.}$ \\
\hline
 & & & $1s^2 2s^1 2p^1$ & 7.4302\\
$^{\ast}Be$&${1s^1 2s^0 2p^3}$&8.9272&${1s^1 2s^0 2p^3}$&8.8051\\
 & & & $1s^1 2s^1 2p^2$& 8.8601\\
\hline
 & & &$1s^1 2s^1 2p^2$ &  6.1351\\
$ Be $&${1s^2 2s^1 2p^1}$&8.9237&${1s^2 2s^1 2p^1}$&8.9217\\
\hline
 & & &$1s^2 2s^2 2p^1$ & 15.0828\\
$ B $ &${1s^2 2s^0 2p^3}$&15.1482&${1s^2 2s^0 2p^3}$&15.1471\\
 &  & &$1s^2 2s^1 2p^2$& 15.1365\\
\hline
 & & &$1s^2 2s^2 2p^2$ & 22.9556\\
$ C $ &  ${1s^2 2s^1 2p^3}$&22.9759&${1s^2 2s^1 2p^3}$&22.9723\\
 &  & &$1s^2 2s^0 2p^4$& 22.9473\\
\hline
 & & &$1s^2 2s^2 2p^3$ & 32.6682\\
$ N $ &  ${1s^2 2s^1 2p^4}$&32.6951&${1s^2 2s^1 2p^4}$&32.6949\\
 &  & &$1s^2 2s^0 2p^5$& 32.6712\\
\hline
 & & &$1s^2 2s^2 2p^4$ & 43.2215\\
$ O $ &  ${1s^2 2s^0 2p^6}$&43.3618&${1s^2 2s^0 2p^6}$&43.3608\\
 & & &$1s^2 2s^1 2p^5$ & 43.3252\\
\hline
 & & &$1s^1 2s^1 2p^6 3s^2$& 55.4295\\
$ F $ &${1s^2 2s^2 2p^3 3s^2}$&55.8686&${1s^2 2s^2 2p^3 3s^2}$&55.6436\\
\hline
 & & &$1s^1 2s^2 2p^6 3s^1$& 69.2101\\
$ Ne $ &${1s^1 2s^1 2p^6 3s^2}$&70.1743&${1s^1 2s^1 2p^6 3s^2}$&69.4399\\
\hline

\end{tabular}
\end{table}

\begin{table}
\caption{Caption is the same as in Table~{IV}.}
\vspace{0.2in}
\begin{tabular}{|c|c|c|c|c|}
\hline
atoms/ions & $exact~config.$ & $-\langle H_0\rangle_{exact}~{\bf a.u.}$&$alt.~config.$&
$-\langle H_0\rangle_{alt.}~{\bf a.u.}$ \\
\hline
 & & &${1s^1 2s^1 2p^2}$& 16.6106\\
$^{\ast}B^{+}$&${1s^1 2s^0 2p^3}$&16.6626&$1s^1 2s^0 2p^3$&16.5501\\
 & & &$1s^2 2s^1 2p^1$& 14.2959\\
\hline
 & & &$1s^1 2s^1 2p^2$& 92.9276\\
$Ne^{6+}$&${1s^1 2s^0 2p^3}$&93.0672&${1s^1 2s^0 2p^3}$&92.9699\\
\hline
 & & &$1s^2 2s^2 2p^6 3s^2 3p^0$& 110.9091\\
$Mg$&${1s^2 2s^2 2p^6 3s^0 3p^2}$&110.9422&${1s^2 2s^2 2p^6 3s^0 3p^2}$&110.9368\\
 & & &$1s^2 2s^2 2p^6 3s^1 3p^1$& 110.9341\\
\hline
 & & &$1s^2 2s^2 2p^6 3s^1 3p^2$ &137.1856\\
$^{\ast}Al$&${1s^2 2s^2 2p^6 3s^0 3p^3}$&137.1905&${1s^2 2s^2 2p^6 3s^0 3p^3}$&137.1853\\
 & & &$1s^2 2s^2 2p^6 3s^2 3p^1$& 137.1655\\
\hline
 & & &$1s^2 2s^2 2p^6 3s^2 3p^2$& 166.2411\\
$^{\ast}Si$&${1s^2 2s^2 2p^6 3s^1 3p^3}$&166.2441&${1s^2 2s^2 2p^6 3s^1 3p^3}$&166.2393\\
 & & &$1s^2 2s^2 2p^6 3s^0 3p^4$& 166.2201\\
\hline
 & & &$1s^2 2s^2 2p^6 3s^2 3p^1$& 170.5714\\
$Si^{+}$&${1s^2 2s^2 2p^6 3s^0 3p^3}$&170.5966&${1s^2 2s^2 2p^6 3s^0 3p^3}$&170.5941\\
 & & &$1s^2 2s^2 2p^6 3s^1 3p^2$& 170.5923\\
\hline
 & & &$1s^2 2s^2 2p^6 3s^2 3p^2$& 203.2693\\
$^{\ast}P^{+}$&${1s^2 2s^2 2p^6 3s^1 3p^3}$&203.2722&${1s^2 2s^2 2p^6 3s^1 3p^3}$&203.2687\\
 & & &$1s^2 2s^2 2p^6 3s^0 3p^4$& 203.2522\\
\hline
 & & &$1s^2 2s^2 2p^6 3s^2 3p^3$& 198.0818\\
$P$&${1s^2 2s^2 2p^6 3s^0 3p^5}$&198.1074&${1s^2 2s^2 2p^6 3s^0 3p^5}$&198.1058\\
 & & &$1s^2 2s^2 2p^6 3s^1 3p^4$& 198.1025\\
\hline
 & & &$1s^2 2s^2 2p^6 3s^2 3p^4$& 232.0761\\
$S$&${1s^2 2s^2 2p^6 3s^0 3p^6}$&232.1008&${1s^2 2s^2 2p^6 3s^0 3p^6}$&232.0996\\
 & & &$1s^2 2s^2 2p^6 3s^1 3p^5$& 232.0957\\
\hline

\end{tabular}
\end{table}

\begin{table}
\caption{Shown in the table atoms/ions with the original excited-state configuration in the
second column and density of this generated by various alternative configurations in the
fourth column.In the third  and  fifth column given are the values of the kinetic energies
corresponding to the original and alternative configurations respectively}
\vspace{0.2in}
\begin{tabular}{|c|c|c|c|c|}
\hline
atoms/ions &$true~config.$&$T[\rho]~{\bf a.u.}$&$alt.~config.$&$T_s[\rho]~{\bf a.u.}$\\
\hline
& & &$1s^1 2s^1 2p^2$& 10.0177\\
$Be$&${1s^1 2s^0 2p^3}$&10.1489&$1s^1 2s^0 2p^3$&10.1481\\
& & &${1s^2 2s^1 2p^1}$& 8.1357\\
\hline
& & &$1s^2 2s^2 2p^1$& 23.7627\\
$B$&${1s^2 2s^0 2p^3}$&24.1249&${1s^2 2s^0 2p^3}$&24.1211\\
& & &$1s^2 2s^1 2p^2$& 23.9238\\
\hline
& & &$1s^2 2s^2 2p^2$& 37.2985\\
$C$&${1s^2 2s^1 2p^3}$&37.5938&${1s^2 2s^1 2p^3}$&37.5922\\
& & &$1s^2 2s^0 2p^4$& 37.9299\\
\hline
& & &$1s^2 2s^0 2p^5$& 30.5856\\
$N$&${1s^1 2s^0 2p^6}$&38.5551&${1s^1 2s^0 2p^6}$&38.5525\\
& & &$1s^2 2s^1 2p^4$& 30.6238\\
\hline
& & &$1s^2 2s^1 2p^5$& 44.6244\\
$O$&${1s^1 2s^1 2p^6}$&54.7136&${1s^1 2s^1 2p^6}$&54.7095\\
& & &$1s^2 2s^0 2p^6$& 44.5899\\
\hline
& & &$1s^2 2s^2 2p^5$& 97.8733\\
$F$&${1s^2 2s^1 2p^6}$&98.5267&${1s^2 2s^1 2p^6}$&98.5212\\
\hline
& & &${1s^2 2s^2 2p^4 3s^1}$& 97.8746\\
$F$&${1s^2 2s^2 2p^3 3s^2}$&98.2631&$1s^2 2s^2 2p^3 3s^2$&98.2393\\
\hline
& & &${1s^1 2s^2 2p^6 3s^1}$& 93.4337\\
$Ne$&${1s^1 2s^1 2p^6 3s^2}$&94.6521&$1s^1 2s^1 2p^6 3s^2$&94.6364\\
\hline
\end{tabular}
\end{table}

\begin{table}
\caption{Caption is the same as in Table~{VI}.}
\vspace{0.2in}
\begin{tabular}{|c|c|c|c|c|}
\hline
atoms/ions &$true~config.$&$T[\rho]~{\bf a.u.}$&$alt.~config.$&$T_s[\rho]~{\bf a.u.}$\\
\hline
& & &${1s^1 2s^1 2p^2}$& 16.6581\\
$B^+$&${1s^1 2s^0 2p^3}$&16.8390&$1s^1 2s^0 2p^3$&16.8378\\
& & &$1s^2 2s^1 2p^1$& 13.2701\\
\hline
 & & &${1s^1 2s^1 2p^2}$& 76.3318\\
$Ne^{6+}$&${1s^1 2s^0 2p^3}$&76.5893&$1s^1 2s^0 2p^3$&76.5837\\
\hline
& & &$1s^2 2s^2 2p^6 3s^1 3p^1$& 199.2404\\
$Mg$&${1s^2 2s^2 2p^6 3s^0 3p^2}$&199.3771&${1s^2 2s^2 2p^6 3s^0 3p^2}$&199.3661\\
& & &$1s^2 2s^2 2p^6 3s^2 3p^0$& 199.1455\\
\hline
 & & &$1s^2 2s^2 2p^6 3s^1 3p^2$& 241.3098\\
$Al$&${1s^2 2s^2 2p^6 3s^0 3p^3}$&241.5112&${1s^2 2s^2 2p^6 3s^0 3p^3}$&241.4967\\
 & & &$1s^2 2s^2 2p^6 3s^2 3p^1$& 241.1428\\
\hline
 & & &$1s^2 2s^2 2p^6 3s^2 3p^2$& 288.4802\\
$Si$&${1s^2 2s^2 2p^6 3s^1 3p^3}$&288.7507&${1s^2 2s^2 2p^6 3s^1 3p^3}$&288.7335\\
 & & &$1s^2 2s^2 2p^6 3s^0 3p^4$& 288.9977\\
\hline
 & & &$1s^2 2s^2 2p^6 3s^1 3p^2$& 287.7594\\
$Si^{+}$&${1s^2 2s^2 2p^6 3s^0 3p^3}$&288.0463&${1s^2 2s^2 2p^6 3s^0 3p^3}$&288.0259\\
 & & &$1s^2 2s^2 2p^6 3s^2 3p^1$& 287.5146\\
\hline
 & & &$1s^2 2s^2 2p^6 3s^2 3p^2$& 339.8495\\
$P^{+}$&${1s^2 2s^2 2p^6 3s^1 3p^3}$&340.1993&${1s^2 2s^2 2p^6 3s^1 3p^3}$&340.1796\\
 & & &$1s^2 2s^2 2p^6 3s^0 3p^4$& 340.5202\\
\hline
 & & &$1s^2 2s^2 2p^6 3s^1 3p^4$& 339.5241\\
$P$&${1s^2 2s^2 2p^6 3s^0 3p^5}$&339.8574&${1s^2 2s^2 2p^6 3s^0 3p^5}$&339.8390\\
 & & &$1s^2 2s^2 2p^6 3s^2 3p^3$& 339.2355\\
\hline
\end{tabular}
\end{table}

\end{document}